\documentclass[footinbib,a4paper,aps,prx,superscriptaddress,reprint,twocolumn,preprintnumbers,amsmath,amssymb,nobalancelastpage,10pt]{revtex4-1}
\usepackage{amsmath}
\usepackage{verbatim}
\usepackage{graphicx}
\usepackage{color}
\usepackage{tikz}
\usepackage{subfigure}
\usepackage{float}

\usepackage{mathptmx}
\usepackage{amssymb}
\usepackage{amsmath}
\usepackage{amsfonts}
\usepackage{array}

\usepackage{bm}
\usepackage{xcolor}
\graphicspath{{figures/}}

\usepackage{braket}

\begin{document}

\title[ATI]{Haake-Lewenstein-Wilkens approach to spin-glasses revisited}

\author{Maciej Lewenstein}
\email[]{maciej.lewenstein@icfo.es}
\affiliation{ICFO - Institut de Ciencies Fotoniques, The Barcelona Institute of Science and Technology, 08860 Castelldefels (Barcelona), Spain}
\affiliation{ICREA, Pg. Llu\'{\i}s Companys 23, 08010 Barcelona, Spain}

\author{David Cirauqui}
\affiliation{ICFO - Institut de Ciencies Fotoniques, The Barcelona Institute of Science and Technology, 08860 Castelldefels (Barcelona), Spain}
\affiliation{Quside Technologies S.L., Carrer d'Esteve Terradas 1, 08860 Castelldefels, Spain}

\author{Miguel Ángel García-March}
\affiliation{Dpto. de Matemática Aplicada,
Universitat Politècnica de València,
Camino de Vera, s/n,
46022 Valencia, Spain}

\author{Guillem Guigó i Corominas}
\affiliation{ICFO - Institut de Ciencies Fotoniques, The Barcelona Institute of Science and Technology, 08860 Castelldefels (Barcelona), Spain}

\author{Przemys\l aw Grzybowski}
\affiliation{Institute of Spintronics and Quantum Information, Faculty of Physics, Adam Mickiewicz University, Umu\l towska 85, 61-614 Pozna\'n, Poland}

\author{José R.M. Saavedra}
\affiliation{Quside Technologies S.L., Carrer d'Esteve Terradas 1, 08860 Castelldefels, Spain}

\author{Martin Wilkens}
\affiliation{Universität Potsdam
Institut für Physik, Karl-Liebknecht-Strasse 24/25,
14476 Potsdam-Golm, Germany}

\author{Jan Wehr}
\affiliation{Department of Mathematics, The University of Arizona Tucson, AZ 85721-0089 USA},

\date{\today}
\pacs{}

\begin{abstract}

We revisit the Haake-Lewenstein-Wilkens (HLW) approach to Edwards-Anderson (EA) model of Ising spin glass [Phys. Rev. Lett. {\bf 55}, 2606 (1985)]. This approach consists in evaluation and analysis of the  probability distribution of configurations of two replicas of the system, averaged over quenched disorder. This probability distribution generates squares of thermal copies of spin variables from the two copies of the systems, averaged over disorder, that is the terms that enter the standard definition of the original EA order parameter,  $q_{\rm EA}$. We use saddle point/steepest descent method to calculate the average of the Gaussian disorder in higher dimensions. This approximate result suggest that $q_{\rm EA}> 0$  at $0<T<T_c$ in 3D and 4D. The case of 2D seems to be a little more subtle, since in the present approach energy increase for a domain wall competes with boundary/edge effects more strongly in 2D; still our approach predicts spin glass order at sufficiently low temperature. We speculate, how these predictions confirm/contradict widely spread opinions that: i) There exist only one (up to the spin flip) ground state in EA model in 2D, 3D and 4D; ii) There is (no) spin glass transition in 3D and 4D (2D). This paper is dedicated to the memories of Fritz Haake and Marek Cieplak.

\end{abstract}

\maketitle
\section{Introduction}

\noindent{\it Spin glass problem.} Spin glasses (SG) have entered solid state and statistical mechanics in  the 1970s, and from the very beginning were considered to be one of the most outstanding and challenging problems of classical statistical physics and theory of disordered and complex systems \cite{Mezard, Chuck}, not to mention their quantum version (cf. \cite{Sachdev,Veronica,Toby} and references therein). The most important and elaborated models of spin glasses are: the Edwards-Anderson (EA) model with short range interactions \cite{EA}, and  the Sherrington-Kirkpatrick (SK) model with infinite range interactions \cite{SK}.\\

\noindent{\it Sherrington-Kirkpatrick model.} The SK model was solved approximately by its inventors using replica trick and replica symmetric solution of the equations that "minimise" the free energy. This solution was clearly physically incorrect, leading to negative entropy at low temperature. G. Parisi found an ingenious way to break the replica symmetry in a hierarchical way \cite{Parisi}. Parisi's solution of the SK model turned out to be exact, first as a local extremum of the free energy \cite{Dedominicis}, and proven rigorously to be unique \cite{Talagrand}. To deepen the understanding of this amazing results it is also recommended to consult the Ref. \cite{Panchenko}. Parisi's solution and Parisi's order parameter, interpreted in terms of probability of overlaps between different frozen configurations of the spin glass is nowadays accepted commonly. For this achievement, and many others, G. Parisi was awarded the Nobel Prize in physics in 2021 for "the discovery of the interplay of disorder and fluctuations in physical systems from atomic to planetary scales."\\

\noindent{\it Edwards-Anderson model.} In the case of EA model, we are very far from a rigorous solution. Most of our knowledge is based on numerical simulations on special purpose classical computers, going back to the 1980s \cite{BY,Ogielski}. It is widely believed that for Ising EA model there is no SG transition at non-zero temperature in 2D, but there is in 3D and higher dimensions. It is not clear that Parisi's picture applies in these low dimensions;  an alternative is provided by the "droplet model" of Ref. \cite{Huse}, which predicts that there exist only one (up to the spin flip) ground state in EA model in 2D, 3D and 4D, but the domain walls, separating the flipped region, from not flipped one, are complex and might even have fractal dimension. Of course, there are some rigorous results concerning EA model (cf. \cite{Arguin}), but they are rather weak and very scarce. Thus, the question of the nature of the SG, as well as many other questions concerning the EA model, is open (cf. \cite{Chuck} and references therein). In recent years various aspects have been studied: ground states in $J=\pm 1$ model \cite{Laundry}, information  theory approach to 3D EA models \cite{people1}, absence of Almeida-Thouless line in 3D SG \cite{Katz1}, or universality in such systems \cite{Katz2}, and several other. The goal of this paper is to look at EA model more than 25 years after the publication of \cite{HLW}, revising the approach developed there.\\

\noindent{\it HLW approach} In 1984 Universität Essen GHS initiated the extremely successful Sonderforschung Bereich "Unordnung und gro\ss e Fluktuazionen" with several neighbouring centres. Fritz was a speaker of this initiative for the next 12 years. He convinced Maciej Lewenstein (his summer-time postdoc) and his new PhD student Martin Wilkens to study short-range spin glasses. They formulated a new approach to this problem, based on idea of studying disorder-averaged probability distribution for configurations of two replicas/copies of the system \cite{HLW}. The idea of the HLW approach is as follows.  We consider two replicas/copies of the system and evaluate the joint probability distribution of configurations averaged over the disorder:
\begin{eqnarray}
    P(\sigma, \sigma')&=&\left\langle\left\langle \exp\{-\beta\left[H(\sigma, \{K_{ij}\}_{\langle ij\rangle}) \right. \nonumber\right.\right.\\
    &+&\left.\left.\left. H(\sigma', \{K_{ij}\}_{\langle ij\rangle)})\right]\}/Z(K)^2\right\rangle\right\rangle, \nonumber 
\end{eqnarray}
where $H(\sigma, \{K_{ij}\}_{\langle ij\rangle})= -\sum_{<ij>} K_{ij}\sigma_i\sigma_j$, $<ij>$ denotes nearest neighbors, $\langle\langle \cdot \rangle \rangle $ denotes average over disorder, and $Z(K)$ is the partition function calculated for a given configuration of the quenched disorder variables $K_{ij}$. We assume that $K_{ij}$ are iidrv's (independent identically distributed random variables) with a Gaussian distribution, $P(K)=\exp(-K^2/2\Delta^2)/\sqrt{2\pi\Delta^2}$ or a binary distribution,  $P(K=\pm \Delta)= 1/2$. Note that both distributions are even, that is, invariant under the change of sign of $K_{ij}$.
The idea is to absorb the sign of $\sigma_i\sigma_j$ into $K_{ij}\to K_{ij}\sigma_i\sigma_j$, and introduce the spin overlap variables $\tau_i=\sigma_i\sigma_i'$. We obtain the effective probability distribution for $\tau$'s
\begin{equation}
    P(\tau)= 2^N\left\langle\left\langle \exp{[\beta \sum_{\langle ij\rangle}K_{ij}(1 + \tau_i\tau_j)]}/Z(K)^2 \right\rangle\right\rangle.
    \label{taureplicas}
\end{equation}
Here the number of relevant variables is reduced as we summed over dummy variables.
Note that magnetic order for $\tau$'s implies the non-zero EA order parameter $q_{\rm EA}$ and vice versa,  $$\langle\sum_i\tau_i\rangle_T/N=\langle\langle\sum_i\langle\sigma_i\sigma_i'\rangle_T/N\rangle\rangle=\sum_i\langle\langle\, \langle\sigma_i\rangle_T^2\rangle\rangle = q_{\rm EA}.$$
We term $\langle \cdot  \rangle $ or $\langle \cdot  \rangle_T $ the thermal average over possible configurations.
Denoting $\alpha=\beta\Delta$, with $\Delta$ the parameter characterizing the probability distributions for the disorder,  HLW used a convenient high temperature expansion to calculate (\ref{taureplicas}) up to 12 order in the expansion parameters $\alpha^2/(1+\alpha^2)$ for the Gaussian, and $\tanh^2(\alpha)/(1 +\tanh^2(\alpha))$ for binary  case. In effect, they calculated
\begin{equation}
    P(\tau)= \exp[- H_{\rm eff} (\alpha, \tau)] /Z_{\rm eff},
    \label{Heff}
\end{equation}
where effective Hamiltonian  contained nearest neighbors couplings $-K_1$, next nearest neighbors couplings $K_2$, and elementary plaquette terms, $K_3$. The coefficients of these terms were explicit functions of
temperature ($\alpha$ in the notation of the present paper). The critical surface separating ferromagnetic from paramagnetic region was estimated  then using (optimized) real space renormalization group approach. It turned out that in
2D the $H_{\rm eff}$ never enters the ferromagnetic region, in 4D  it enters the ferromagnetic region for sure, and in 3D the situation was not clear, suggesting that $H_{\rm eff}$ touches the critical region in a quadratic manner. That would imply that the critical exponents of the spin glass model are two times bigger than those of the standard Ising model, in agreement with the best numerical simulation available at that time.\\

\noindent{\it HLW followers} The paper by HLW did not found too many followers, but some very prominent are worth mentioning. Indeed, R. Swendsen with collaborators published two papers on HLW method in Phys. Rev. B in the end of 1980s. In the first one by J.-S. Wang and R.H. Swendsen \cite{Swendsen1}, the authors studied Monte Carlo renormalization-group  of Ising spin glasses. Application of
this approach to the $\pm J$ Ising spin glass showed clear differences between 2D, 3D, and 4D
models. The data were consistent with a zero-temperature transition in two dimensions,
and non-zero temperature transitions in three and four dimensions. In another paper  \cite{Swendsen2} Monte Carlo and high-temperature-expansion calculations
of a spin-glass effective Hamiltonian were performed. The authors studied the quenched random-coupling spin-glass problem from the point of view of a nonrandom
effective Hamiltonian, by Monte Carlo and high-temperature-expansion methods. It was found
that the high-temperature series of the  spin-glass effective Hamiltonian diverges below
the ferromagnetic transition temperature. The Monte Carlo approach does give reliable results at
low temperatures. The results were compared with the HLW picture of
spin-glass phase transitions.\\

\noindent{\it Present work.} In this paper we revise HLW approach. The idea is to estimate $P(\tau)$, performing saddle point/steepest descent approximation in calculating the Gaussian average of the disorder, which should be correct in the limit $\alpha
\to \infty$. We argue that the resulting spin model has couplings that are positive in the region where $\tau_i\tau_j=1$'s, so it has tendency to order ferromagnetically on islands/domains, separated from other domains by negative couplings. In effect,   boundary/edge effects start to play a role in estimates of various quantities that may characterize the order in our system.

We present here various arguments in favor or against the spin glass order (ferromagnetic order in overlap variables).
First, we consider the original Peierls' argument \cite{Peierls, Griffiths}, and argue that in our situation, it can hardly be used. We turn then to an argument, studying sensitivity of the system to boundary conditions.  This argument was originally proposed by Thouless \cite{Thouless1,Thouless2,gang} for models of electron propagation in the presence of disorder and subsequently adapted to study Ising models in random magnetic fields \cite{ImryMa} (see also \cite{Chudnovsky}), and also spin glasses \cite{Cieplak}.
This argument is relating the existence of the ferromagnetic phase transition to the sensitivity to boundary conditions. It can be trivially used for ferromagnetic spin models: it "predicts" transition for $d\ge 2$ for Ising model, no transitions for $d=2$ models with continuous symmetry (Mermin-Wagner-Hohenberg theorem), and transitions for $d\ge 3$ for systems with continuous symmetry, like $XY$ or Heisenberg models (cf. \cite{LSA17}).  

To apply this argument,  we calculate $P_+=P(\tau_i=+1,\, _{\rm for\ all\ {\it i}'s})$ on a cylinder in $d$ dimensions of cross-section $L^{d-1}$ and length $L$, and compare it to $P_-=P(\{\tau\}={\rm corresponding\ to\ one\ domain\ wall})$. We analyze $\delta = \log(P_+/P_-)$ and argue that this quantity, within approximations used,  is always positive and proportional to $L^{d-1}$ in $d\ge 2$. We will argue that the situation in 2D seems to be a little more complex because of the stronger interplay between the boundary effects and the domain wall energy. This leads to significantly higher critical temperature in 2D than in higher dimensions.
 
\section{Saddle point/steepest descent calculations}

We focus here on the case of Gaussian disorder, since we are going to use differential calculus.
First, we rescale $K_{ij}=\alpha\Delta \kappa_{ij}$, so that both the logarithm of the distribution of $\kappa_{ij}$, and the logarithm of $P(\tau)$ become proportional to $\alpha^2$ as $\alpha^2\to\infty$. The HLW formula becomes
\begin{equation}
    P(\tau)= 2^N\left\langle\left\langle \exp{[\alpha^2 \sum_{\langle ij\rangle}\kappa_{ij}(1+\tau_i\tau_j)]}/Z(\kappa)^2 \right\rangle\right\rangle ,
    \label{taurrenors}
\end{equation}
where the average $\langle\langle \cdot \rangle \rangle$ is now with respect  the distribution $P(\kappa)=\exp(-\alpha^2\kappa^2/2)/\sqrt{2\pi/\alpha^2}$.\\

\noindent{\it Laplace's method.} The idea is to calculate the asymptotic behavior of the disorder average using the Laplace method, also known as the saddle point/steepest descent (SPSD) method, which we expect to be asymptotically accurate for $\alpha\to\infty$. The SPSD equations equating to zero the first derivatives of the logarithm of the integrand with respect to $\kappa_{ij}$'s read:
\begin{equation}
0=\alpha^2(-\kappa_{ij} + 1+\tau_i\tau_j-2\langle \sigma_i\sigma_j\rangle),
\end{equation}
where $\langle \sigma_i\sigma_j\rangle$ is the thermal average of the neighboring spins correlator, calculated according to the canonical distribution
$P(\sigma)=\exp{[\alpha^2\sum_{\langle ij\rangle}\kappa_{ij}\sigma_i\sigma_j]}/Z(\alpha^2 k)$. There are two possibilities:
\begin{itemize}
\item $\tau_i\tau_j=1$. In this case:
\begin{equation}
\kappa_{ij}=2(1-\langle \sigma_i\sigma_j\rangle)>0,
\end{equation}
so that the corresponding coupling is clearly ferromagnetic.

\item $\tau_i\tau_j=-1$. In this case
\begin{equation}
\kappa_{ij}=-2\langle \sigma_i\sigma_j\rangle,
\end{equation}
and the situation is more delicate. For $\alpha$ large, if $\langle \sigma_i\sigma_j\rangle)>0$, we expect the coupling $\kappa_{ij}$ to be ferromagnetic, but the above equation implies the opposite.  Likewise, if the correlation function is negative, the $\kappa_{ij}<0$ should be ferromagnetic. The contradiction could be avoided if $\kappa_{ij}=0$, but the true situation is more complex, as we will see below, by solving systematically mean field equations. This contradiction is really an expression of frustration in our system!
\end{itemize}

It follows that we can write the SPSD solutions as $\kappa_{ij}>0$ on the domains, where neighboring $\tau_i\tau_j=1$. This solution has a very clear meaning: the canonical ensemble that serves to calculate the correlation functions $\langle \sigma_i\sigma_j\rangle$ corresponds to ferromagnetic islands/domains (where $\tau_i\tau_j=1$), separated by domain walls, where the bonds $\kappa_{ij}\le 0$, $\tau_i\tau_j=-1$, and the correlations between $\sigma$'s from different domain walls  are still positive, but perhaps smaller at the border. 

Note that the situation we consider is not as in the standard spin glass, where we look at $\langle \sigma_i\sigma_j\rangle$ for a fixed configuration of random $\kappa_{ij}$'s. There, it is quite common that the sign of $\kappa_{ij}$ is not equal to the sign of  $\langle \sigma_i\sigma_j\rangle$: this is actually how the frustration exhibits itself basically! Here, however, we consider a different situation: for a given configuration of $\tau$'s, we adjust the values of $\kappa_{ij}$'s to satisfy the SPSD equations.  The natural expectation is  a ferromagnetic order for $\tau$'s (i.e. SG order for $\sigma$'s) in our system, with the energy (free energy/probability) cost of the energy wall to scale as $L^{d-1}$, as in, say, the standard Ising ferromagnet. At the same time, we cannot exclude the existence of other solutions of SPSD equations that would inherit frustration more explicitly. We discuss this possibility, which goes beyond the scope of the present paper, in the outlook.\\
 
\noindent{\it Hessian matrix.} In the zeroth order one can calculate now $P(\tau)$, substituting for $\kappa_{ij}$'s their SPSD values. One can go one step further calculating the Gaussian correction to the SPSD. To this aim we calculate the Hessian matrix of the second derivatives of the logarithm of the integrand. Let us introduce the shortened notation $(ij)=\mu$, $(i'j')=\nu$, $\sigma_i\sigma_j= c_\mu$, $\sigma_{i'}\sigma_{j'}= c_\nu$, etc. The Hessian matrix reads
\begin{equation}
{\cal H}_{\mu\nu} =-\alpha^2[\delta_{\mu\nu} + \alpha^2[\langle c_\mu c_\nu\rangle-\langle c_\mu\rangle\langle c_\nu\rangle]].
\end {equation}
Note that the correlations matrix
$$ \langle c_\mu c_\nu\rangle-\langle c_\mu\rangle\langle c_\nu\rangle=\langle (c_\mu -\langle c_\mu\rangle)(c_\nu -\langle c_\nu\rangle)\rangle,$$
i.e. it is explicitly positively semi-definite. In effect the Hessian matrix:
\begin{equation}
\hat {\cal H} <0,
\end{equation}
so that the logarithm of the integrated function, which we consider is a strictly convex function of many variables, is expected to have one maximum, corresponding to our SPSD solutions. Note also that eigenvalues of the Hessian matrix are all negative and will typically be of order $\alpha^4$, and they are bounded in modulus from below by $\alpha^2$. One should thus expect that SPSD method should become  for $\alpha\to \infty$ asymptotically very precise, if not exact.

\section{Peierls and Thouless approaches}
\label{PT estimates}

In this section we examine if the $\tau$ variables of our effective model for two copies of the EA systems exhibite ferromagnetic order i.e. if the EA order parameter is nonzero, signifying spin glass order. We present two approaches: i) Peierls approach; ii) Thouless approach; in the latter case we first discuss several analytic estimates, and then present self-consistent calculations, using SPSD solutions for $\kappa_{ij}$'s as a point of departure for local mean field calculations of the averages of $\tau$'s and $\tau-\tau$ correlations.\\

\noindent{\it Peierls approach.} Peierls considers domain walls in a square lattice in 2D, defining them in  an unambiguous way. In a ferromagnetic Ising model with the uniform coupling $K$ (with $\beta$ absorbed into $K$), and  with periodic boundary conditions on a square of  side $L$, and number of sites $N=L^2$,  with all spins $\tau_i=1$ on the boundary, all domain walls are closed. Let $b$ denote the length of the domain's boundary;  We classify them according to length $b$, and
within a class of given length we give each a number $i$. A wall of the length $b$ fits into a square of the side $b/4$ and area $b^2/16$. Let $m( b)$ be the number of domain walls of length $b$; it is obviously bounded by $m(b) \le 4N 3^{b-1}$. The next step is to consider the quantity 
$X(b,i)=1$, if the domain wall $(b, i)$ occurs in that configuration, and $X(b,i) = 0$ otherwise. Clearly, 
the number of spins down fulfills: 
$$N_-/N\le \sum_b(b^2/16)3^b\sum_i^{m(b)}X(b,i).$$
Peierls estimates  then the thermal average of $X(b,i)$ in the Gibbs-Boltzmann ensemble, bounding the partition function from below by the contribution of the configuration, in which all spins inside the considered domain were flipped, and obtaining the bound  $\langle X(b,i)\rangle \le \exp[-2\beta K b]$, which leads to the desired result. Notably, it can be generalized to higher dimensions, with a little extra effort to estimate the entropy of contours, see \cite{Bonati} for an elementary discussion and references therein for original work.

Unfortunately, we cannot use this reasoning, because in our case: i) couplings are non-homogeneous; ii) their values depend on domain walls configurations, according  to SPSD equations.  We can estimate that the configuration  $C$, in which the domain $(b,i)$ occurs, has contribution to the "energy" coming from two edges, $4\alpha^2\kappa_{\rm e}$, where $\kappa_{\rm e}$ is the coupling on the edge. The configuration $\tilde C$, in which the spins inside the wall are flipped, contributes to $Z(\kappa)$ with the energy larger by $3\alpha^2\kappa$, with $\kappa$ being the coupling in the bulk, so that 
$\langle X(b,i)\rangle \le \exp\{\alpha^2 [4\kappa_{\rm e}-6 \kappa] b\}$. Since, according to mean field, $\kappa_{\rm e}> \kappa$, the question is to be able to estimate more precisely the interplay of the edge and bulk contributions. To this aim we turn, however, to a simpler Thouless argument, to decide about the existence of the magnetization, i.e. spin glass order.\\

\noindent{\it Thouless argument.} In order to investigate the sensitivity to boundary conditions, we calculate $P_+=P(\{\tau_i=1\}_{\rm for\ all\ i's})$ on a cylinder in $d$ dimensions of cross-section $L^{d-1}$ and length $L$, and compare it to $P_-=P(\{\tau_i=1\}_{\rm on\ the\ left}, \{\tau_i=-1\}_{\rm on\ the\ right})$ with $\tau_i=\pm 1$ on the left (right ) of a domain wall (DW), correspondingly. We determine the parameter $\delta=\ln(P_+/P_-)$; Ferromagnetic order for $\tau$'s (SG order for $\sigma$'s) is indicated by $\delta>0$

We consider a lattice with coordination number $f$, with $f_{\rm out}$ bonds sticking out at any site  of any $L^{d-1}$-dimensional hyper-plane (cross-section). As we will see, we will need to compare the effects of DW and boundary effects, since both scale as $L^{d-1}$. To this end we will also consider effective coordination number at the edge (boundary) hyper-planes, $f_{\rm e}$.
Geometrically, $f_{\rm e}=f-f_{\rm out}/2$ for the left and right edge (boundary) hyper-planes -- only half of $f_{\rm out}$ stick out to the right (left) from the left (right) edge. This is evidently a good estimate for $f_{\rm e}$ in higher dimensions, where we expect $f_{\rm out}\ll f$. On the other hand, boundary effects do extend to more than just the edge, so it is reasonable to approximate
\begin{equation}
f_{\rm e}\le f-f_{\rm out}.
\end{equation}

We stress that we are NOT considering here domain walls in the disordered EA model, where they are believed to a have a very complex geometry, scaling and maybe even effective dimension, in accordance with the seminal droplet model \cite{Huse}. We are studying here domain walls in  the effective, averaged over disorder, probability distribution of the $\tau$ variables. Just from the construction, there are no reasons  for this probability distribution to break translation symmetry (everywhere, i.e. in $d$ dimensions,  if we apply global periodic boundary conditions, or at least in $(d-1)$ transverse dimensions, if we apply periodic boundary conditions there). It is thus natural to look in the first place for domain walls that are just flat hyper-planes.\\

    \section{Self-consistent SPSD and local MF solutions}
    
In the following we focus on hyper-cubic lattices in $d$-dimensions, with coordination number $f=2d$, $f_{\rm out}=2$, and $f_{\rm e}\le 2d-1$. We leave the preliminary discussion of other lattices to appendix C and future publication. 
In this section, we estimate the bulk and the edge contributions applying SPSD and local MF consistently from the beginning till the end. We consider a $d$-dimensional cylinder of spins with $L$ layers with bonds distributed according to a Gaussian distribution $P(K) = \exp(-K^2/2\Delta^2)/\sqrt{2\pi\Delta^2}$, at an inverse temperature $\beta$. We denote as above $\alpha = \Delta\beta$. \\

\noindent{\it Local mean field theory.} We assume translation symmetry in $d-1$ transverse dimensions, so that magnetisation depend only on one index, $i$, enumerating the layer, and the couplings depend on two indices, enumerating involved single layer (two neighboring layers). Using standard mean field theory (MF), we find the magnetization that is the thermal average of $\langle \sigma_i\rangle$ at the  $i$-th layer as
\begin{eqnarray} \label{eq: mi = Fi(m)}
    m_i &=&\tanh\left[ \alpha^2 \left( 2\left(d-1\right)\kappa_{i,i}m_i   \right.\right.\\
    &+& \left.\left.\kappa_{i,i+1}m_{i+1} + \kappa_{i-1, i}m_{i-1}\right) \right] , \nonumber 
\end{eqnarray}
with 
\begin{equation}
    \kappa_{i,j} = 2 - 2m_i m_j,
\end{equation}
and boundary conditions:
\begin{equation} \label{eq: boundary condition m}
    m_0 = m_{L+1} = 0,
\end{equation}
\begin{equation} \label{eq: boundary condition k}
    \kappa_{0, 1} = \kappa_{L, L+1} = 0.
\end{equation}\\

\noindent{\it Quantities to be determined.} Our aim is to calculate logarithm of the probability $P_+$, $P_-$ and $\delta= \ln(P_+/P_-)$. We denote $\ln(P_\pm)=H_\pm$, and call it "energy" in the following, so that
\begin{equation}
    H_+ = L\ln(2)+\alpha^2 \sum_{i, j} [2\kappa_{i, j}-\kappa_{i, j}^2/2] - 2\ln Z(\kappa),
\end{equation}
Being an extensive quantity, the energy of the system divided by the volume in all but one dimension is
\begin{eqnarray} \label{eq: H_+ / L^d-1}
    \frac{H_+}{L^{d-1}} &=& L\ln(2)+ \alpha^2 \left[ \left(d-1\right)\sum_{i = 1}^{L}[2\kappa_{i, i}-\kappa_{i, i}^2/2] \right.\\
    &+& \left.\sum_{i=1}^{L-1}[2\kappa_{i, i+1}-\kappa_{i, i+1}^2/2] \right] - \sum_{i=1}^{L}2\ln\left[2\cosh(F_i(m))\right]\nonumber\\
    &-&\frac{1}{2}\ln({\rm det}(\hat H_+)).\nonumber
\end{eqnarray}
with $F_i\left(m\right) =\alpha^2(2(d-1)\kappa_{i,i} m_i +\kappa_{i-1,i}m_{i-1}  + \kappa_{i,i+1}m_{i+1})$. Note that we have included in this expression the term coming from the Gaussian fluctuations around the SPSD solution. The above  quantity in the leading order should be a linear function of the cylinder's length, $$H_+/L^{d-1} = A(\alpha)L + B_+(\alpha).$$ 

A similar expression holds for $P_-$, also including Gaussian fluctuations terms: 
\begin{eqnarray} \label{eq: H_- / L^d-1}
    \frac{H_-}{L^{d-1}} &=& L\ln(2)+ \alpha^2 \left[ \left(d-1\right)\sum_{i = 1}^{L}[2\kappa_{i, i}-\kappa_{i, i}^2/2] \right.\\
    &+& \sum_{i=1}^{L/2-1}[2\kappa_{i, i+1}-\kappa_{i, i+1}^2/2]  +   \sum_{i=L/2+1}^{L-1}[2\kappa_{i, i+1}-\kappa_{i, i+1}^2/2] \nonumber\\
    &-& \left.\kappa_{L/2, L/2+1}^2/2\right]- \sum_{i=1}^{L}2\ln\left[2\cosh(F_i(m))\right].\nonumber\\
    &-&\frac{1}{2}\ln({\rm det}(\hat H_-)).\nonumber
\end{eqnarray}
Since configuration contributing to $P_-$ has connection between two layers in the middle of the cylinder given by a different expression, clearly $$H_-/L^{d-1} = A(\alpha)L + B_-(\alpha),$$ with the same bulk contribution, but different boundary term;   thus
$$\delta=B_+(\alpha)-B_-(\alpha).$$
Positive value of $\delta$ indicates ferromagnetic order for $\tau$'s and spin glass order for $\sigma$'s. \\

To calculate $H_-/L^{d-1}$ we repeat the above calculations using the same formulae as before, except that we use
\begin{equation}
    \kappa_{L/2,L/2+1} = - 2m_{L/2} m_{L/2+1},
\end{equation}. \\

\noindent{\it Gaussian fluctuation terms} Generally speaking, Gaussian fluctuation terms play a sub-leading role, as expected. We approximate
$\ln({\rm det}(\hat H_\pm))=\sum_\mu \ln(\lambda_\mu)\approx \sum_\mu \ln(\hat H_{\mu\mu})$, that is the sum of logarithms of eigenvalues by the sum of logarithms of diagonal elements of the Hessian matrix. Noting that 
$$
\partial \langle \sigma_i \sigma_j\rangle/\partial \kappa_{ij}=\alpha^2
\left(1-m_i^2m_j^2\right),$$
we obtain
\begin{equation}
\label{Gaussian}
-\frac{1}{2}\ln({\rm det}(\hat H_\pm))\approx -\frac{1}{2} \sum_{(ij)}\ln[\alpha^2(1+ \alpha^2 (1-m_i^2m_j^2)], 
\end{equation}
where the SPSD solutions for $m$'s are calculated for the case $\pm$ accordingly. The above expression undergoes, obviously,  further simplifications under the translation symmetry.\\

\noindent{\it High $\alpha$ regime} Before going to numerical solutions, we first analyze the asymptotic regime $\alpha\to \infty$, where $A(\alpha)$ can be estimated analytically.
We consider MF equations in the bulk of the $d$-dimensional hyper-cubic lattice. 
The corresponding self-consistent equations in the bulk are:
\begin{eqnarray}
\kappa&=&2(1-m^2), \\
m&=& \tanh(2\alpha^2 d\kappa  m), \\
g&=&\ln(2\cosh(2\alpha^2 d\kappa  m).
\end{eqnarray}
We transform the first two into an equation of $x=2\alpha^2 d \kappa$.
\begin{equation}
x= 4\alpha^2 d/\cosh^2(x \sqrt{1-\kappa/(4\alpha^2 d2)}).
\end{equation}
For large $\alpha$ we get $x=\ln(4\alpha^2 d)/2$, and $\kappa=\ln(4\alpha^2 d)/(4\alpha^2 d)$. As expected,
$\kappa\to 0$ as $\alpha\to \infty$, and $m\to 1$, but $2\alpha^2 d \kappa$ diverges as $\ln(4\alpha^2 d)$. 
Elementary analysis leads to the result:
$$A(\alpha)\simeq \ln(2)-\frac{1}{2}\ln(4\alpha^2 d),$$
i.e. as expected $\ln(P_\pm)=H_\pm$ becomes negative at large $L$ (when our analysis makes sense) and at large 
$\alpha$ (when SPSD should work well); $A(\alpha)$ diverges with $\alpha$, but very slowly, only logarithmically. 

In calculation of asymptotic behavior of $\alpha^2\kappa$. we typically set local magnetization to 1: they indeed tend to one, but in slightly different way in the bulk and on the ends, as the numeric illustrates below. If we set $m_i=1$ in Eq. (\ref{eq: H_+ / L^d-1}), and expand for large $\alpha$, then we obtain a simple expression for
\begin{eqnarray} \label{eq: H asymt}
    \frac{H_+}{L^{d-1}} &=& L\ln(2)-2\alpha^2 \left[ \left(d-1\right)\sum_{i = 1}^{L}\kappa_{i, i} \right.\\
    &+& \left.\sum_{i=1}^{L-1}\kappa_{i, i+1} \right],\nonumber
\end{eqnarray}
neclecting sub-leading Gaussian corrections. Since our numerical analysis in the asymptotic regime is tough, we may and will use this expression there. The analysis is more complex in the case of $\frac{H_-}{L^{d-1}}$, where we need to take into account the dramatic change of the nature of SPSD solutions at the domain wall. \\

\noindent{\it "Phase transition" at moderate $\alpha$} The solution of the MF equations change character as $\alpha$ grows from small values (when all $m_i=0$) to larger values (when all $m_i\ne 0$). We  infer the existence of this "phase transition" at a finite $\alpha$ by imposing that solutions get trivial at that point, $\alpha_T$. This way we can approximate Eq. (\ref{eq: mi = Fi(m)}) for temperatures close to $\alpha_T$ as a series expansion for small $m_i$ to get:
\begin{equation}
    m_i \approx \alpha^2\left( 2\left(d-1\right)\kappa_im_i + \kappa_{i-1, i}m_{i-1} \kappa_{i+1}m_{i+1}\right)
\end{equation}
To first order, $\kappa_{i, j}=2$ and $m_i=m$ $\forall i$, so we find the critical temperature:
\begin{equation}
    \alpha_T = \frac{1}{2\sqrt{d}}
\end{equation}\\

\noindent{\it Numerical calculations} By numerically solving the system of equations $F(m) = m_i$ for $1\leq i\leq L$ and taking into account that in positions $i = \{0, L+1\}$ there are no spins and therefore conditions Eq. (\ref{eq: boundary condition m}) and Eq. (\ref{eq: boundary condition k}) apply, we find non-trivial solutions above a certain temperature threshold, see Fig. \ref{fig: mi, d = 2}.
\begin{figure}[ht]
\centering
\includegraphics[scale=.5]{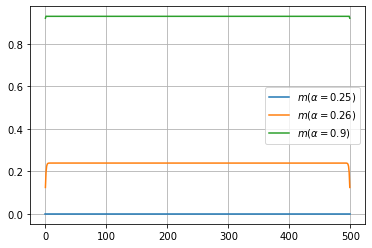}
\includegraphics[scale=.5]{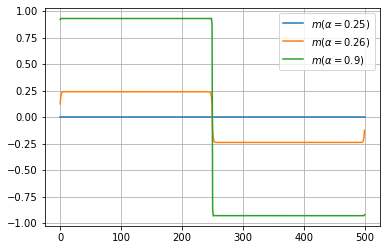}
\caption{Numerical solutions of the system of Eqs. (\ref{eq: mi = Fi(m)}) for $d=4$, $L=500$ at various temperatures, corresponding to the case of $P_+$ (upper panel) and  $P_-$ (lower panel. In the latter case, the solutions change the sign of $m$'s in the middle (and the keep the sign of $\kappa$'s positive). Solutions for $d$ equal to 2 and 3 are qualitatively the same, and quantitatively very similar.}
\label{fig: mi, d = 2}
\end{figure}

\begin{figure}[ht]
\centering
\includegraphics[scale=.5]{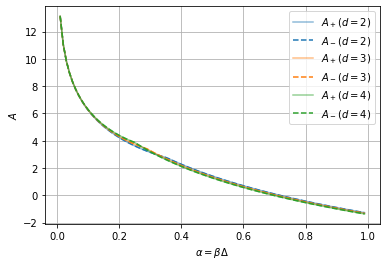}
\includegraphics[scale=.5]{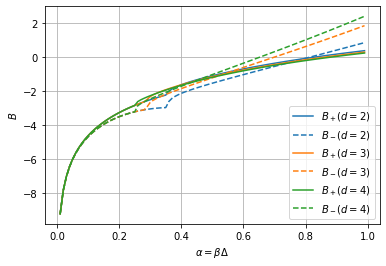}
\caption{Numerical solutions for $A$ (upper panel) and $B_+$ and $B_-$ (lower panel) for $d$ equal 2, 3, 4, and at various temperatures. Gaussian fluctuations contributions are included.}
\label{fig: AandB}
\end{figure}

We solve the system of equations for various lengths $L$ and fit the obtained results in order to obtain $A$ and $B$ at different temperatures, Fig. \ref{fig: AandB}. We do so for dimensions $d = 2, 3, 4$ and obtain similar behaviours. As expected, MF solutions for all three systems undergo a "phase transition" from $m=0$ to $m\ne 0$ at their respective critical temperatures,  $\alpha_T^{d = (2, 3, 4)} = \{ \frac{1}{2\sqrt{2}} , \frac{1}{2\sqrt{3}}, \frac{1}{4} \}$.

\begin{figure}[ht]
\centering
\includegraphics[scale=.5]{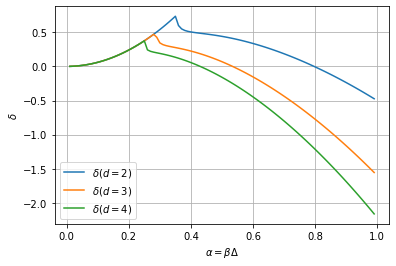}
\caption{The parameter $\delta$ for $d = 2, 3, 4$ as a function of $\alpha$ (temperature). Gaussian fluctuations contribution is included.}
\label{fig: delta}
\end{figure}

 The results show in accordance with analytic calculations that $A(\alpha)$  tends to $-\infty$ logarithmically. On the other hand, $B_+(\alpha)$ tend to a positive constant for large $\alpha$,  while $B_-(\alpha)$ to infinity, indicating SG transition in $2D$ (unfortunately), $3D$ (fortunately), and $4D$ (fortunately). This is illustrated clearly in Fig. \ref{fig: delta}, Still, one observes quite a quantitative difference in behaviour for $d=2$ and $d$ larger. 

\section{Conclusions and outlook}

In the short note we revised the Haake-Lewenstein-Wilkens (HLW) approach to Edwards-Anderson (EA) model of Ising spin glass. The main results are the following:
\begin{itemize}
    \item We have calculated the disorder averaged probability of spin configurations for two replicas, which reduces to a probability of overlaps between spins from the two replicas, $P(\tau)$. To this aim we used the saddle point/steepest descent (SPSD) method which seems to be asymptotically exact in the limit of $\alpha=\beta\Delta$ going to infinity.The integral we consider, has an integrand, whose logarithm has a well peaked single maximum, with the Hessian of order at least $\alpha^2$, if not $\alpha$. It would be challenging to study if one can control this result rigorously.
    \item We attempted  to apply Peierls and Thouless approaches to decide whether there exist SG order in the low temperature (large $\alpha^2$ limit). The results indicate that this indeed is the case in 2D and above, but we identified the reasons, why this does not have to be the case in 2D. Namely, the competing boundary effects might destroy the order. Our estimates, based on mean field theory, clearly require improvement, for instance by studying precisely  the solutions of boundary effects in SPSD equations etc. If we accept the proposed form of the solutions of the SPSD solutions, the simulating $P_+$ requires MC simulations of a finite size ferromagnetic model, while simulating $P_-$ -- also a finite size ferromagnetic model with a domain wall and a bump/dip in the couplings at the wall.
    \item In a nutshell: Our results predict SG transition in EA model in $4d$, $3d$, but unfortunately also in $2d$. There can be several reasons for that: i) SPSD approximation is not precise enough; ii) is completely incorrect; In the first case we can include Gaussian and maybe even beyond Gaussian corrections to SPSD solutions. In the second case, there might be many SPSD solutions contributing or something like that; Hessian result suggests this is not the case, but it is not rigorous; iii) finally, local MF calculations of edge/boundary effects might be too rough.  
    \item The paper contains 4 appendices: In Appendix A we  discuss shortly the exactly soluble 1D case, in Appendix B -- the normalization of $P(\tau)$ that implies nice properties of certain multidimensional integrals. Of course, the present results are compatible with the expectation that there exist only one (up to the spin flip) ground state in EA model in 2D and 3D \cite{Laundry}. Another interesting conclusion is that the existences of the SG transition in the present picture, might depend on the connectivity  of the lattice. As discussed in Appendix \ref{AppendixC}, even within our SPSD and MF domain walls have a certain width. This might depend crucially on the dimension and even on the coordination number (connectivity) of the lattice. Finally, alternative way of calculations combining SPSD method with the expected behavior of $Z(\kappa)$ for large $\alpha$ is discussed in Appendix D. This method explicitly accounts for dependence of correlators $\langle\sigma_i\sigma_j\rangle$ on $\kappa_{ij}$,
\end{itemize}

Clearly, this study requires further studies, but this goes beyond the present note. \\

\acknowledgments{ICFO group acknowledges support from ERC AdG NOQIA, State Research Agency AEI (“Severo Ochoa” Center of Excellence CEX2019-000910-S) Plan National FIDEUA PID2019-106901GB-I00 project funded by MCIN/ AEI /10.13039/501100011033, FPI, QUANTERA MAQS PCI2019-111828-2 project funded by MCIN/AEI /10.13039/501100011033, Proyectos de I+D+I “Retos Colaboración” RTC2019-007196-7 project QUSPIN funded by MCIN/AEI /10.13039/501100011033, Fundació Privada Cellex, Fundació Mir-Puig, Generalitat de Catalunya (AGAUR Grant No. 2017 SGR 1341, CERCA program, QuantumCAT \ U16-011424, co-funded by ERDF Operational Program of Catalonia 2014-2020), EU Horizon 2020 FET-OPEN OPTOLogic (Grant No 899794), and the National Science Centre, Poland (Symfonia Grant No. 2016/20/W/ST4/00314), Marie Sk\l odowska-Curie grant STREDCH No 101029393, “La Caixa” Junior Leaders fellowships (ID100010434),  and EU Horizon 2020 under Marie Sk\l odowska-Curie grant agreement No 847648 (LCF/BQ/PI19/11690013, LCF/BQ/PI20/11760031,  LCF/BQ/PR20/11770012).}


\appendix
\section{Exact solution 1D}

Calculation of $P(\tau)$ in 1D are elementary. We observe first that 
$$Z(\kappa)=\prod_{i=1}^{i=L-1}2\cosh(\alpha^2\kappa_{i,i+1}),$$
so that $P(\tau)$ can be written as
\begin{eqnarray}
    P(\tau)&=&\langle\langle\frac{2^L}{Z^2(\kappa)}\prod_{i=1}^{i=L-1}\left[\cosh(\alpha^2\kappa_{i,i+1}) + \sinh(\alpha^2\kappa_{i,i+1})\right]\nonumber\\
    &&\left[\cosh(\alpha^2\kappa_{i,i+1}) + \tau_i\tau_{i+1}\sinh(\alpha^2\kappa_{i,i+1})\right]\rangle\rangle.
\end{eqnarray}
Since we average over the even distributions the terms $\cosh(.)\sinh(.)$ average zero, and we get 
\begin{equation}
    P(\tau)=2\prod_{i=1}^{i=L-1}\left[1 + \tau_i\tau_{i+1}\langle\langle\tanh^2(\alpha^2\kappa)\rangle\rangle\right],
\end{equation}
where we skipped the subscript of $\kappa$. We can again  estimate $\langle\langle\tanh^2(\alpha^2\kappa)\rangle\rangle$ using SPSD. Saddle point value for $2\alpha^2\kappa$ diverges again as $\ln(4\alpha^2)$, so the 1D system exhibits a "phase transition" at zero temperature ($\alpha \to\infty$) with diverging correlation length $\xi\propto \alpha^2$.

\section{Normalization issues - amazing formulae}

Note that if we observe that $P(\tau)$, by definition is normalized
\begin{equation}
    P(\tau)= 2^N\left\langle\left\langle \exp{[\beta \sum_{\langle ij\rangle}K_{ij}(1 + \tau_i\tau_j)]}/Z(K)^2 \right\rangle\right\rangle  , 
    \label{taureplicasB}
\end{equation}
then by tracing over $\tau$'s we obtain
\begin{equation}
    1= 2^N\left\langle\left\langle \exp{[\beta \sum_{\langle ij\rangle}K_{ij}]}/Z(K) \right\rangle\right\rangle .
    \label{taureplicasB1}
\end{equation}
The above expression is true for any even distribution of $K$'s, Gaussian or not, discrete or continuous. It can be generalized to certain matrix models with couplings invariant with respect to local unitary transformations. The independent proof of this formula employs the fact that
$$2^N = \sum_{\sigma}1.$$
Using the above formula and then incorporating each of the configurations of $\sigma$'s into the averaging over disorder, gives the desired identity. 

\section{Domain wall width}
\label{AppendixC}

\begin{figure}[ht]
\centering
\includegraphics[scale=.5]{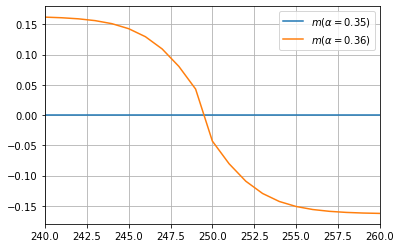}
\includegraphics[scale=.5]{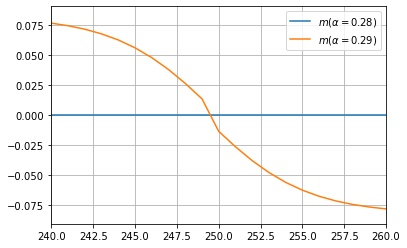}
\includegraphics[scale=.5]{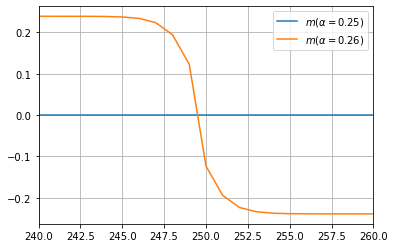}
\caption{Numerical solutions for magnetization in the domain wall region for $d=2$ (upper panel), $d=3$ (middle panel), and  for $d=4$ (lower panel) at indicated temperatures.}
\label{magn}
\end{figure}

It is worth noticing the domain walls in the case of $P_-$ have a finite width. This means that local magnetization $m_i$ does not jump from nearly one to nearly minus one (see Fig. \ref{magn}). In effect, $\kappa$'s in the domain wall regions are not so close to zero, and the terms $\alpha^2\kappa$ simply behave in this region as $\alpha^2$. This explain the rapid growth of $B_-$ in Fig. \ref{fig: AandB}.

Sharpening of the domain wall to the configuration that $m=\pm 1$ in the bulk, and $m=0$ at the domain wall edges would, would lead presumably to instability of the ferromagnetic phases. In fact we have originally postulated (incorrect) solutions of SPSD equations with $\kappa=0$ at the walls. Such solution leads to $B_-=2B_+$ -- it still predicts the ferromagnetic order, but with very different, much more milder behavior of $\delta$. Conversely, widening the wall, more in the spirit of the "droplet model" might also lead to unexpected behavior, since the assumption that $\ln(P_{\pm 1}) = AL + B_{\pm 1}$ would then cease to hold. 

Our numerical findings with the SPSD and MF approximations indicate that: i) for fixed $\alpha^2$, the domain wall reaches an $L$-independent limit for $L$ large; ii) for fixed $L$, the domain wall shinks from $L$ (below phase transtion, where all $m$'s are zero), to a very small values dictated by the very fast growth of $|m|$'s toward one, in accordance with the MF laws.

\section{Alternative approach}
\label{AppendixD}

Here we propose alternative way of calculating $\delta$ based on expected behaviour of $\ln(Z(K))$ for low temperatures. Namely, we expect that 
$$- 2\ln (Z(K))=2\beta F\simeq 2 \beta \langle U \rangle, $$
where $U$ is the internal energy. That means that in the SPSD method we need to analyse the logarithm of the integral kernel:
\begin{equation}
\alpha^2 \sum_{\langle ij\rangle}[\kappa_{ij}(1+\tau_i\tau_j- 2\langle \sigma_i\sigma_j\rangle) -\kappa_{ij}^2/2 ]
\end{equation}
The equations for $\kappa$'s are modified due to the explicit dependence of $\langle \sigma_k\sigma_l\rangle$ on $\kappa_{ij}$; in fact one easily gets
$$
\partial \langle \sigma_k \sigma_l\rangle/\partial \kappa_{ij}=\alpha^2
\left(\langle \sigma_k \sigma_l\sigma_i\sigma_j\rangle- \langle \sigma_k\sigma_l\rangle\langle \sigma_i\sigma_j\rangle\right).$$
Fortunately, most of these correlators are negligible: in fact they vanish in the MF approximation for distinct, non-overlaping pairs $(k,l)$ and $(i,j)$. The non-vanishing and non-trivial are 
$$
\partial \langle \sigma_i \sigma_j\rangle/\partial \kappa_{ij}=\alpha^2
\left(1-m_i^2m_j^2\right),$$
and 
$$
\partial \langle \sigma_i \sigma_l\rangle/\partial \kappa_{ij}=\alpha^2
\left(m_lm_j(1-m_i^2)\right), $$
and its variations.
We obtain then modified equations for $\kappa_{ij}$ that have now to be solved in an iterative manner, 
$$\kappa_{ij} = \frac{1+\tau_i\tau_j-2m_im_j -2\Delta \kappa_{ij} } {1+ 2 \alpha^2(1- m_i^2 m_j^2)}, $$
where 
$$\Delta \kappa_{ij}= \alpha^2\left(\sum_{l={\rm n.n.}} \kappa_{il}m_lm_j(1-m_i^2) + \sum_{k={\rm n.n.}} \kappa_{kj}m_im_k(1-m_j^2)\right),$$
where $l$'s ($k$'s) and neighbors of $i$ ($j$), different from $j$ ($i$). 

These expressions get simplified upon translation symmetry, 
\begin{eqnarray}
\Delta \kappa_{i,i}&=& 2\alpha^2\left(\kappa_{i,i}(2d-3) m_i^2(1-m_i^2) \right.\\
&+&\left. m_i(\kappa_{i,i+1}m_{i+1}+\kappa_{i-1,i} m_{i-1})(1-m_{i}^2)  \right),\nonumber
\end{eqnarray}
\begin{eqnarray}
\Delta \kappa_{i,i+1}&=& \alpha^2\left(2(d-1) \kappa_{i,i+1}m^2_{i+1}(1-m_i^2) \right. \\
&+&  2(d-1) \kappa_{i,i+1} m_i^2(1-m_{i+1}^2) \nonumber\\
 &+& \left.\kappa_{i-1,i}m_{i-1}m_{i+1}(1-m_{i}^2) +
 \kappa_{i,i+1}m_{i}m_{i+2}(1-m_{i+1}^2)\right).\nonumber
\end{eqnarray}

For $P_+$ and $P_-$ (away from the wall) we get
$$\kappa_{ij} = \frac{2-2m_i m_j-2\Delta \kappa_{ij}} {1+ 2 \alpha^2(1- m_i^2 m_j^2)},$$
and at the wall
$$\kappa_{L/2,L/2+1} = \frac{-2m_{L/2} m_{L/2+1}-2\Delta \kappa_{L/2,L/2+1}} {1+ 2 \alpha^2(1-m_{L/2}^2m_{L/2+1}^2)}.$$
Otherwise, all other expressions are valid. The calculations of $P_+$ and $P_-$ reduces now to evaluation of 
\begin{eqnarray} \label{eq: H_+ / L^d-1new}
    \frac{H_+}{L^{d-1}} &=& L\ln(2)+ \alpha^2 \left[ \left(d-1\right)\sum_{i = 1}^{L}[2\kappa_{i, i}(1-m_i^2)-\kappa_{i, i}^2/2] \right.\\
    &+& \left.\sum_{i=1}^{L-1}[2\kappa_{i, i+1}(1-m_im_{i+1})-\kappa_{i, i+1}^2/2] \right] .\nonumber
\end{eqnarray}
and 
\begin{eqnarray}
 H_-/L^{d-1}&=&\alpha^2 \sum'_{\langle ij\rangle}[2\kappa_{ij} -\kappa_{ij}^2/2 - 2\kappa_{ij} m_i m_j]\\ \nonumber
&+&\alpha^2[-\kappa_{L/2,L/2+1}^2/2 - 2\kappa_{L/2,L/2+1} m_{L/2} m_{L/2+1}].
\end{eqnarray}
Similarly, 
\begin{eqnarray} \label{eq: H_- / L^d-1new}
    \frac{H_-}{L^{d-1}} &=& L\ln(2)+ \alpha^2 \left[ \left(d-1\right)\sum_{i = 1}^{L}[2\kappa_{i, i}(1-m_i^2)-\kappa_{i, i}^2/2] \right.\nonumber\\
    &+& \sum_{i=1}^{L/2-1}[2\kappa_{i, i+1}(1-m_im_{i+1})-\kappa_{i, i+1}^2/2] \\
    &+&   \sum_{i=L/2+1}^{L-1}[2\kappa_{i, i+1}(1-m_im_{i+1})-\kappa_{i, i+1}^2/2] \nonumber\\
    &-& \left.\kappa_{L/2, L/2+1}^2/2-\kappa_{L/2, L/2+1}m_{L/2}m_{l/2+1} \right].\nonumber
\end{eqnarray}
We have calculated $\delta$, using the present approach, in which we neglected contributions from $\Delta\kappa$'s terms, leaving only the effect due to $
\partial \langle \sigma_i \sigma_j\rangle/\partial \kappa_{ij}=\alpha^2
\left(1-m_i^2m_j^2\right)$. The end results are quantitatively and qualitatively the very similar to those obtained with the "pure" SPSD method.

\end{document}